\documentclass[a4paper,11pt]{article}

\usepackage{authblk}
\usepackage{amsmath,amssymb,amsfonts}
\usepackage{color}
\usepackage{graphicx}

\newcommand{\be}{\begin{equation}}
\newcommand{\ee}{\end{equation}}

\newcommand{\G}{\mathcal{G}}
\newcommand{\LL}{\mathcal{L}}

\newcommand{\R}{\mathbb{R}}
\newcommand{\rme}{{\rm e}}
\begin{document}

%%%% Article title to be placed here
\title{
The Fokker--Planck equation of the superstatistical fractional 
Brownian motion with application to passive tracers inside cytoplasm}

\author{%%%% Author details
C. Runfola$^{1,2}$, S. Vitali$^{2,3}$ and G.~Pagnini$^{2,4\footnote{Corresponding author: gpagnini@bcamath.org}}$\\
\small
$^{1}$Department of Physics and Astronomy, University
of Bologna, Viale Berti Pichat 6/2, I-40127 Bologna, Italy\\ 
$^{2}$BCAM -- Basque Center for Applied Mathematics,
Alameda de Mazarredo 14, E-48009 Bilbao, Basque Country -- Spain\\ 
$^{3}$Eurecat, 
Centre Tecnol{\'o}gic de Catalunya, 
Unit of Digital Health, Data Analytics in Medicine, Barcelona, 
E-08005, Catalunya -- Spain\\
$^{4}$Ikerbasque -- Basque Foundation for Science,
Plaza Euskadi 5, E-48009 Bilbao, Basque Country -- Spain
}

\maketitle

%%%% Abstract text to be placed here %%%%%%%%%%%%
\begin{abstract}
By collecting from literature data
the experimental evidences of anomalous diffusion 
of passive tracers inside cytoplasm, %living systems,
and in particular of subdiffusion of mRNA molecules inside live E. coli cells,
%[Golding \& Cox, Phys. Rev. Lett. 2006], 
we get the probability density function of molecules' displacement 
and we derive the corresponding Fokker--Planck equation.
Molecules' distribution e\-mer\-ges to be related to the Kr{\"a}tzel function
and its Fokker--Planck equation 
be a fractional diffusion equation in the Erd{\'e}lyi--Kober sense.
The irreducibility of the derived Fokker--Planck equation to 
those of other literature models is also discussed.
\end{abstract}

%%%% Subject entries to be placed here %%%%
\noindent
{\bf Subject:} mathematical physics, statistical physics, biophysics.

%%%% Keyword entries to be placed here %%%%

%%%% Insert corresponding author and its email address}

\section{Introduction}
The experimental evidence of anomalous diffusion in living systems
has been definitively established 
\cite{barkai_etal-pt-2012,
hofling_etal-rpp-2013,
manzo_etal-rpp-2015} and, 
in particular, we remind the measurements 
of the motion of mRNA molecules inside live E. coli cells 
by Golding \& Cox \cite{golding_etal-prl-2006} 
that are now a milestone in the field. 
Here we derive the Fokker--Planck equation for the
probability density function (PDF) of molecules' displacement
in agreement with the main findings from 
such dataset and similars.

Unfortunately, the PDF of passive tracers in cytoplasm 
is not available yet from data because of technical issues that
limit the number of particle's trajectories and this affects, in particular, 
the reliability of the tails of the distribution which are the footprint
of deviation from Gaussianity and standard diffusion.
Thus, in these 
circumstances with lack of measurements,
we do not analyse experimental data
but we collect experimental results 
\cite{golding_etal-prl-2006,
magdziarz_etal-prl-2009,
magdziarz_etal-pre-2010,
burnecki_etal-pre-2010,
weron_etal-prl-2010,
magdziarz_etal-pre-2011,
mackala_etal-pre-2019,
janczura_etal-csf-2022}
in order to derive first 
the molecules' PDF and then the corresponding Fokker--Planck equation.
If the solution, namely the PDF, is already known
then the governing equation can be considered useless but,
in the following, we discuss how indeed knowing the 
PDF is not all in the game and the knowledge of the corresponding 
equation provides indeed futher valuable information.

In general, similar features of anomalous diffusion 
have been experimentally observed in 
the motion of different passive tracers in different
living systems, see, e.g., 
\cite{tolicnorrelykke_etal-prl-2004,
bronstein_etal-prl-2009,weigel_etal-pnas-2011,
tabei_etal-pnas-2013,
regner_etal-bj-2013,jeon_etal-prl-2011,manzo_etal-prx-2015}, 
and the fractional Brownian motion (fBm) 
emerged to be the underlying random motion of molecules' trajectories,
see, e.g., 
\cite{szymanski_etal-prl-2009,
weiss-pre-2013,krapf_etal-prx-2019,
sabri_etal-prl-2020,
han_etal-elife-2020,itto_etal-jrsi-2021,korabel_etal-e-2021},
as well as the generalised Gamma distribution with its
special cases emerged to be the distribution of the diffusion coefficients,
e.g.,
\cite{hapca_etal-jrsi-2009,petrovskii_etal-an-2009,manzo_etal-prx-2015}.
Therefore, 
a superstatistical fBm
resulted to be a successful model 
\cite{molina_etal-pre-2016,mackala_etal-pre-2019,itto_etal-jrsi-2021}.
%namely  the process $X_t=\sqrt{\Lambda} \, \mathcal{B}^H_t$ 
%where $\Lambda$ is a non-negative random variable 
%properly distributed and
%$\mathcal{B}^H_t$ is the fBm with Hurst exponent $H \in (0,1)$,

The superstatistical fBm %$X_t=\sqrt{\Lambda} \, \mathcal{B}^H_t$
is indeed a randomly scaled Gaussian process
that was studied for fractional anomalous diffusion 
originally within the framework of the 
generalised gray Brownian motion (ggBm)  
\cite{mura_etal-jpa-2008,grothaus_etal-jfa-2015,
grothaus_etal-jfa-2016, dasilva_etal-s-2015,
molina_etal-pre-2016}, 
% \,{\buildrel d \over =}\, (L_\alpha)^{-\alpha}$, 
%with $\alpha \in (0,1)$
%and $L_\alpha$ 
that recently has been extended to investigate the relation with
generalised time-fractional diffusion equations
\cite{bender_etal-fcaa-2022a,bender_etal-fcaa-2022b}.
Moreover, 
randomly scaled Gaussian processes resembling the ggBm 
have been formulated 
for modelling and understanding anomalous diffusion
also within an under-damped approach 
\cite{vitali_etal-jrsi-2018,sliusarenko_etal-jpa-2019,chen_etal-njp-2021}.

The qualitative success of the superstatistical fBm for modelling anomalous
diffusion in living systems was already discussed within the
framework of the ggBm in relation to ergodicity breaking 
and fractional diffusion \cite{molina_etal-pre-2016}.
The PDF of molecules
is therefore a mixture of Gaussian distributions 
with random variances and, within a Bayesian approach, 
such population can be understood as the likelihood modulated 
by the prior distribution of a parameter. 
The formal randomization of this parameter  
is equivalent to the computation of the marginal likelihood, 
which corresponds indeed to the PDF of i.i.d. random variables
\cite{vitali_etal-m-2019}. This view allows for highlighting and
clarifing the role of the central limit theorem in 
the dynamics of an heterogeneous ensemble of Brownian particles
as in the superstatistical fBm here considered \cite{vitali_etal-m-2019}.
Any value of the variance of the Gaussian PDFs %$\G(\cdot)$ 
is idiosyncratic and the random diffusion coefficients %variable $\Lambda$ 
represent the heterogeneity of the ensemble of molecules,
which is sufficient to (weakly) break 
the ergodicity of the system \cite{molina_etal-pre-2016}.

To conclude, we report that
the superstatical modelling of anomalous diffusion can be related 
also with the effect of probes' polidispersity that generates
an apparent anomalous diffusion as emerged in the cytoplasm of human cells
where the apparent anomaly exponent decreases 
with increasing polydispersity of the probes
\cite{kalwarczyk_etal-jpcb-2017}. These results can be applied 
also in intracellular studies of the mobility of nanoparticles, 
polymers, or oligomerizing proteins.

Here we found that the PDF of molecule displacement belongs to the 
family of the Kr{\"a}tzel special function \cite{princy-cms-2014}
and its Fokker--Planck equation 
is a fractional diffusion equation in the Erd{\'e}lyi--Kober sense 
\cite{luchko_etal-fcaa-2013,mathai_haubold-2018} 
that cannot be reduced to existing models for anomalous diffusion.

The rest of the paper is organised as follows.
In the next Section \ref{sec:data} we present the dataset
by Golding \& Cox \cite{golding_etal-prl-2006} and the
related findings from the corresponding data analysis.
In Section \ref{sec:PDF} we establish the molecule PDF and
in Section \ref{sec:FPeq} we derive the governing Fokker--Planck equation.
In Section \ref{sec:discussion} we discuss the results and in particular
the valuable information behind the determination of the 
Fokker--Planck equation in spite of the known PDF.
The last Section \ref{sec:conclusions} is devoted to the conclusions
and future perspectives.  

\section{The paradigmatic dataset by Golding \& Cox}
\label{sec:data}
In this study, we are interested in the governing equation of
the PDF of passive tracers diffusing in cytoplasm because of 
the information that it can give
about the process, see \S \ref{sec:discussion}, 
apart from the PDF itself that is its fundamental solution.
In particular, for our aims, 
we mainly consider the results coming from the data acquired 
by Golding \& Cox \cite{golding_etal-prl-2006}.
The trajectories included in that dataset are 21 
and so they are not enough for calculating ensemble averages. 
In this respect,
we remind here that anomalous diffusion is indeed characterised 
also by a non-Gaussian distribution of particles 
and therefore also by the scaling of the tails of the distribution
which are indeed affected by the size of the sample of the observations.

In their experiment, 
they considered the motion of mRNA molecules released 
from their template DNA and free to move in the cytoplasm,
in particular, 
they tracked the random motion of 
individual fluorescently labeled mRNA molecules inside live E. coli cells.
This dataset turned to be a benchmark dataset for studying 
anomalous diffusion in living systems and widely analysed
in these years, see, e.g., 
\cite{magdziarz_etal-prl-2009,
magdziarz_etal-pre-2010,
burnecki_etal-pre-2010,
weron_etal-prl-2010,
magdziarz_etal-pre-2011,
mackala_etal-pre-2019,
janczura_etal-csf-2022}.
The main findings from the Golding \& Cox experiment
\cite{golding_etal-prl-2006} can be summerized as follows. 

Let $\mathcal{X}_t$ be the molecule position at time $t > 0$,
the time-averaged mean square displacement (TA-MSD) of molecules
resulted to be subdiffusive 
$\overline{[\mathcal{X}_{t+\Delta} - \mathcal{X}_t]^2} \sim \Delta^\beta$
\cite{golding_etal-prl-2006}, with $\beta \in (0,1)$.
Moreover, TA-MSD emerged to be quite scattered 
\cite{golding_etal-prl-2006,mackala_etal-pre-2019,janczura_etal-csf-2022} 
and this can be due to a population of diffusion coefficients
that leads to weak ergodicity 
breaking \cite{manzo_etal-prx-2015,molina_etal-pre-2016}.

Furthermore, the application of the method of p-variation
showed that the underlying motion of the molecules is more likely 
the fBm 
\cite{magdziarz_etal-prl-2009,magdziarz_etal-pre-2010,mackala_etal-pre-2019}.
Therefore, by adopting a superstatistics of fBm,
namely by adopting
the process $X_t=\sqrt{\Lambda} \, \mathcal{B}^H_t$ 
where $\Lambda$ is a non-negative random variable and
$\mathcal{B}^H_t$ is the fBm with Hurst exponent $H \in (0,1)$,
it was possible to estimate 
the diffusion coefficients of the molecules' ensemble 
which resulted in a population related to the Weibull distribution 
\cite{mackala_etal-pre-2019},
that actually is a special case of the generalised Gamma distribution. 
The ggBm is recovered when 
$\Lambda$ %\,{\buildrel d \over =}\, (L_\alpha)^{-\alpha}$, 
is a totally skewed positive $\alpha$-stable random variable
with stable index $\alpha \in (0,1)$
\cite{mura_etal-jpa-2008,molina_etal-pre-2016}.
%and $L_\alpha$ is a totally skewed positive $\alpha$-stable random variable,
A Weibull distribution of the diffusion coefficients 
has been confirmed \cite{janczura_etal-csf-2022}
also when the fBm was replaced by the so-called fractional {L\'evy} 
stable motion (FLSM)
that is a L\'evy-driven stochastic process that generalizes the fBm
\cite{burnecki_etal-pre-2010,janczura_etal-csf-2022}.

Trajectories from the Golding--Cox dataset are not enough for
properly testing mixing and ergodicity 
by comparing ensemble and time averages.
However, necessary conditions for mixing and ergodicity
can be derived on the basis of a single trajectory
\cite{magdziarz_etal-pre-2011}.
Trajectories from the Golding--Cox dataset satisfy 
the necessary conditions for mixing and ergodicity 
\cite{magdziarz_etal-pre-2011},
which is consistent with both the fBm 
\cite{mackala_etal-pre-2019} and the 
proper FLSM model \cite{weron_etal-prl-2010,magdziarz_etal-pre-2011}.
Actually, if each trajectory is properly re-scaled 
then their scattering is largely reduced and the  
ergodicity breaking is no longer present \cite{mackala_etal-pre-2019}.

Since the FLSM is based on L\'evy stable distributions,
diverging statistics as, for example, the MSD are replaced
by the sample MSD that may exhibit either normal and anomalous diffusion.
The FLSM results to perform comparable or even better than the fBm,
at least for some of the trajectories, 
for what concerns some observable as the Hurst exponent, 
stability index and both sample MSD and sample p-variation 
\cite{burnecki_etal-pre-2010}.
This may be due to the fact that the experimental trajectories 
may deviate from Gaussianity, contrary to the fBm approach, 
and then the more flexible FLSM allows for catching this feature 
\cite{janczura_etal-csf-2022}.  

Actually, further valuable experimental data 
on diffusion of passive tracers in cytoplasm have been published 
and so their analysis, too, beside those related with 
the Golding--Cox dataset \cite{golding_etal-prl-2006}.
Here, we remind, for example, 
the studies of particle motion in crowded fluids, 
that is an analogue of the motion inside the cytoplasm of living cells, 
by using dextran dissolved in water
\cite{weiss_etal-bj-2004,banks_etal-bj-2005,ernst_etal-sm-2012} 
or purely viscous solution obtained by using sucrose into water
\cite{ernst_etal-sm-2012} 
that led to the understanding of the corresponding diffusive motions
in terms of the fBm \cite{weiss-pre-2013,sabri_etal-prl-2020},
as well as the measurements of the dynamics of 
histonelike nucleoid-structuring proteins in live E. coli bacteria 
\cite{sadoon_etal-pre-2018} where a power-law distribution 
of the diffusion coefficients of individual proteins has been observed
in agreement with the Pearson Type VII distribution that led to the
the development of a modelling approach based on the fBm that hierarchically 
takes into account the joint fluctuations of both 
the anomalous diffusion exponents and the diffusion constants 
\cite{itto_etal-jrsi-2021},
or diffusion of tracer particles in the cytoplasm of mammalian cells
where the experimental observations are described 
by an intermittent fBm alternating between two states 
of different mobility \cite{sabri_etal-prl-2020}
and similarly in the case of
intracellular endosomes \cite{han_etal-elife-2020} suggesting that 
the underlying trajectories can be modelled
by a fBm with a distributed Hurst exponent \cite{korabel_etal-e-2021}.

\section{Superstatistical molecules' distribution}
\label{sec:PDF}
\begin{figure}[!h]
\centering\includegraphics[width=\textwidth]{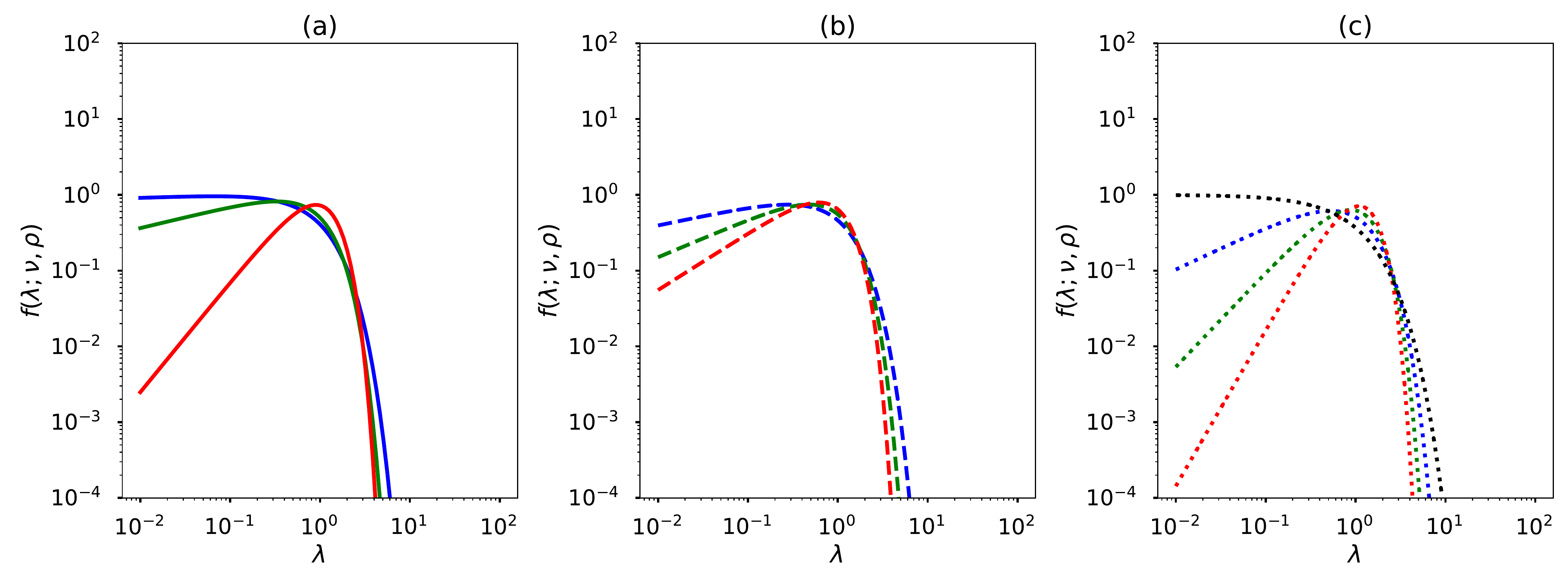}
\caption{Plots of the distribution of the diffusion coefficients
$f(\lambda)$ (\ref{gengamma}) for different values of the
parameters:
the generalised Gamma distribution 
[$\nu=1.25$ and $\rho=1.5$ (solid blue line),
$\nu=1.5$ and $\rho=1.75$ (solid green line),
$\nu=1.75$ and $\rho=1.25$ (solid red line)];
the Weibull distribution 
[$\nu=\rho=1.25$ (dashed blue line),
$\nu=\rho=1.5$ (dashed green line),
$\nu=\rho=1.75$ (dashed red line)];
the Gamma distribution ($\rho=1$)
[$\nu=1.25$ (dotted blue line),
$\nu=1.5$ (dotted green line),
$\nu=1.75$ (dotted red line)];
and the exponential distribution [$\rho=\nu=1$ (dotted black line)].
In all the plots $\lambda_0=1$.
}
\label{fig:lambdadist}
\end{figure}

In the case under consideration,
the PDF of molecules results to be given by the superstatistical integral
\be
P(x;t) = \int_0^\infty 
\G(x;t | \lambda) f(\lambda) d\lambda  \,,
\label{PDFPGf}
\end{equation}
where
\begin{equation}
\G(x; t | \lambda) =
\frac{1}{\sqrt{4\pi \lambda t^{2H}}} \, 
\rme^{-x^2/(4 \lambda t^{2H})} \,,
\label{gaussianfbm}
\end{equation}
is the Guassian PDF of the fBm and $f(\lambda)$ is, in short notation, 
the generalised Gamma distribution %\cite{stacy-1962}
\be
f(\lambda)=f(\lambda ; \lambda_{0}, \nu ,\rho) = 
\frac{\rho }{\lambda_0^\nu \,\Gamma ( \nu/\rho)} \lambda^{\nu - 1} 
\rme^{-( \lambda/\lambda_0)^{\rho} } \,, 
\label{gengamma}
\end{equation}
with $\lambda, \lambda_0, \nu, \rho > 0$. 
The generalised Gamma distribution reduces to 
the Weibull distribution when $\nu=\rho$, i.e.,
$f(\lambda;\lambda_0,\rho,\rho)=W(\lambda;\lambda_0,\rho)$,
to the Gamma distribution when $\rho=1$ and 
to the exponential distribution when $\rho=\nu=1$,
see a comparison in Figure \ref{fig:lambdadist}. 

The generalised Gamma distribution
was studied in the framework of superstatistics 
both in the original formulation \cite{beck-pa-2006} 
and in the recent formulations for anomalous diffusion 
within the diffusing-diffusivity
\cite{sposini_etal-njp-2018,postnikov_etal-njp-2020} 
and the ggBm approaches 
\cite{sposini_etal-njp-2018}. 

\begin{figure}[!h]
\centering\includegraphics[width=\textwidth]{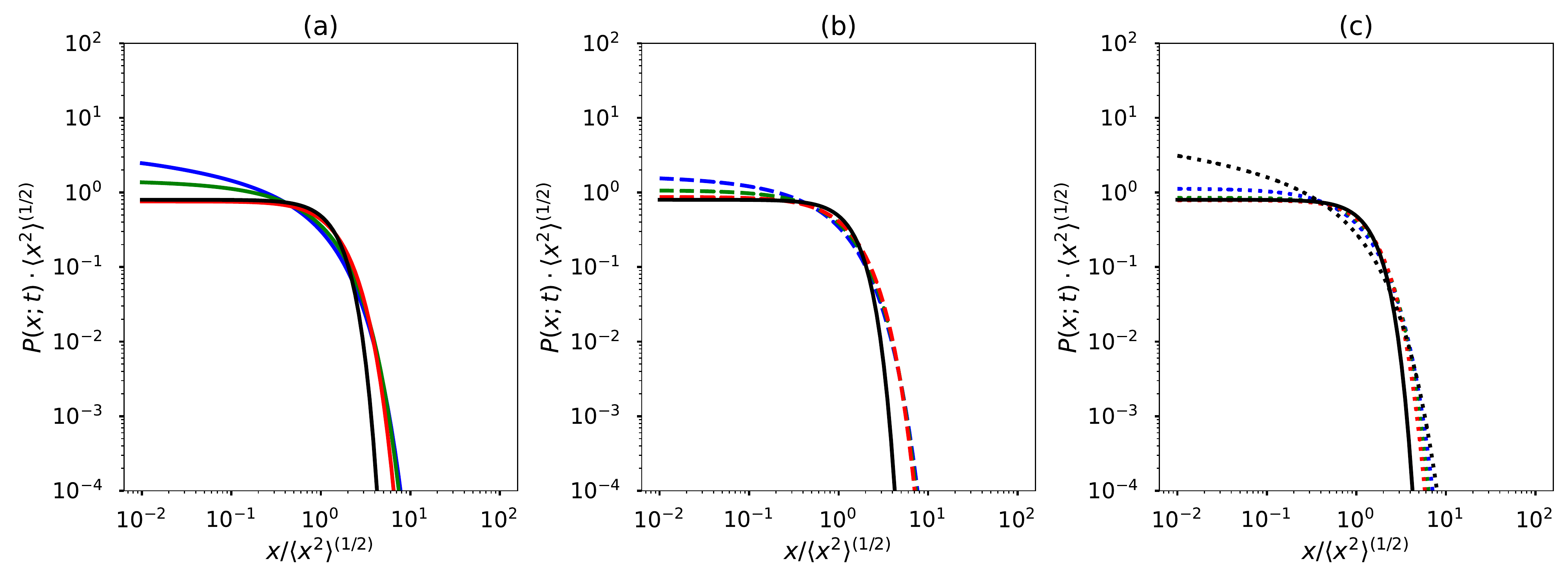}
\centering\includegraphics[width=\textwidth]{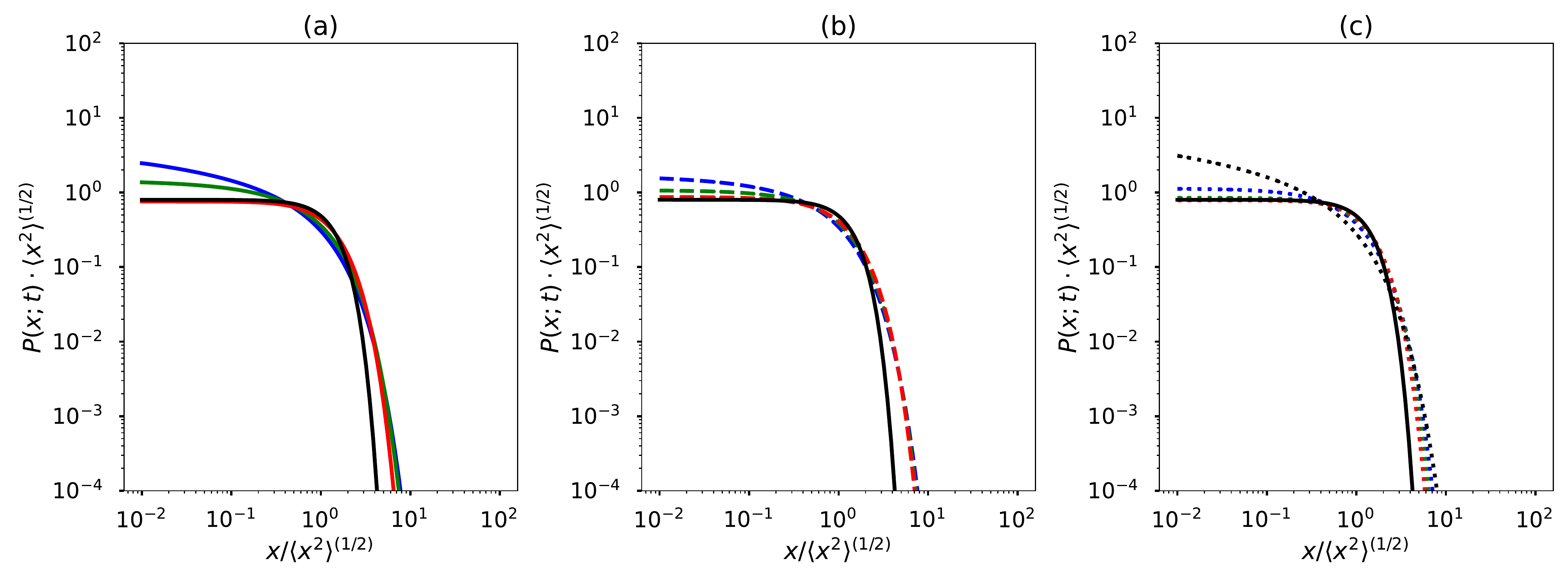}
\centering\includegraphics[width=\textwidth]{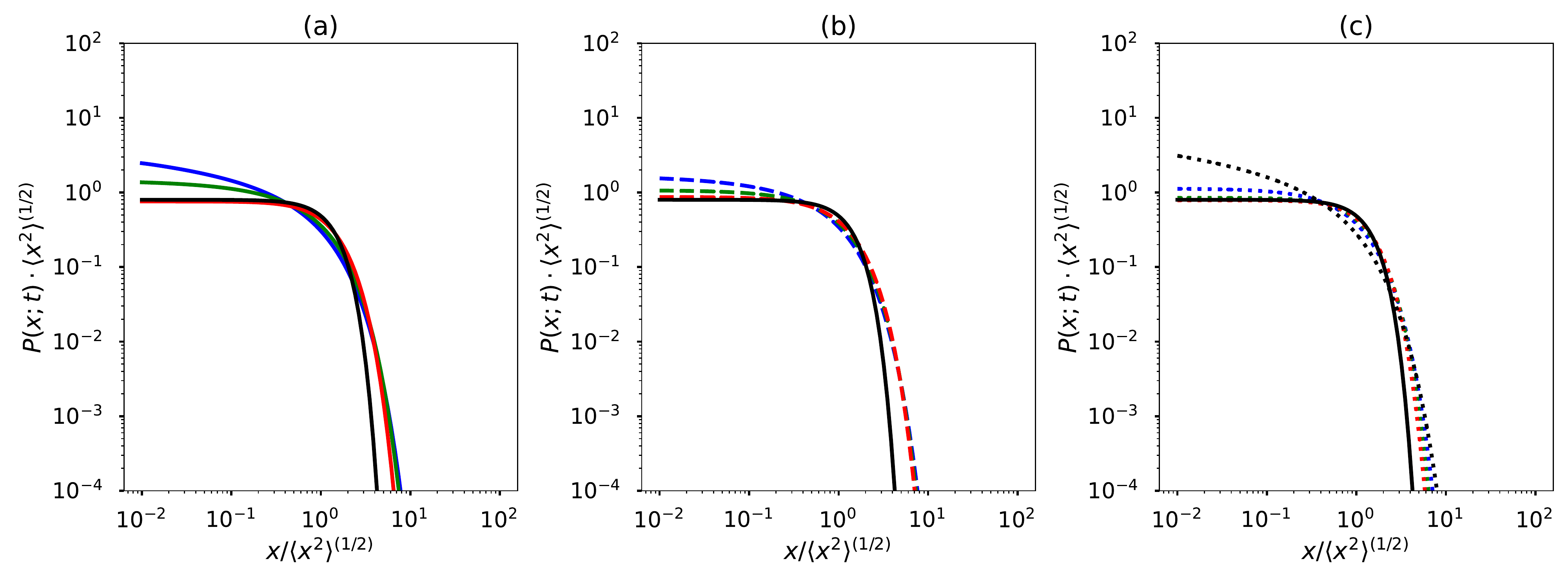}
\caption{Plots of the molecules' PDF (\ref{PDFPGf}),
i.e.,  
$\langle x^2 \rangle^{1/2} P(x;t) \, vs \, x/\langle x^2 \rangle^{1/2}$,
in lin-log scale with the following values of the parameters
of the distribution of the diffusion coefficients $f(\lambda)$ (\ref{gengamma}):
the generalised Gamma distribution 
[$\nu=1.25$ and $\rho=1.5$ (solid blue line),
$\nu=1.5$ and $\rho=1.75$ (solid green line),
$\nu=1.75$ and $\rho=1.25$ (solid red line)];
Weibull distribution with
[$\nu=\rho=1.25$ (dashed blue line),
$\nu=\rho=1.5$ (dashed green line),
$\nu=\rho=1.75$ (dashed red line)];
the Gamma distribution ($\rho=1$)
[$\nu=1.25$ (dotted blue line),
$\nu=1.5$ (dotted green line),
$\nu=1.75$ (dotted red line)];
and the exponential distribution [$\rho=\nu=1$ (dotted black line)].
In all the plots $\lambda_0=1$.
The reference with the Gaussian density is also displayed (solid black line).
Panel a): $H=0.25$; panel b): $H=0.5$; panel c): $H=0.75$;
where any difference due to $H$ is indeed removed
because of the self-similarity with respect to the variance.
}
\label{fig:plotPDF}
\end{figure}

We observe that molecules' PDF (\ref{PDFPGf})
%, which solves the Fokker--Planck equation (\ref{EKeq}),
can be re-written as
\be
P(x;t) = 
\frac{\rho}{\Gamma(\nu/\rho)}
\frac{1}{\sqrt{4 \pi \lambda_{0} t^{2H}}} \, 
Z_\rho^{\nu-\frac{1}{2}} \!
\left(\frac{x^2}{4 \lambda_{0} t^{2H}}\right)
\,,
\label{PDFGamK}
\ee
where $Z_{\rho}^{\nu}(\cdot)$ is the Kr{\"a}tzel function 
\cite{princy-cms-2014}
\be
\displaystyle{
Z_{\rho}^{\nu}(u) = 
\int_0^\infty \lambda^{\nu-1} 
\rme^{ -\frac{u}{\lambda}-\lambda^{\rho}} d\lambda \,, 
\quad u > 0 \,,
}
\end{equation}
and the variance of particle displacement is
\begin{eqnarray}
\langle x^2 \rangle 
&=& 2 \int_0^{\infty} x^2 P(x;t) \, dx \nonumber \\
&=& \int_0^\infty 2 \lambda t^{2H} \, f(\lambda) \, d\lambda \nonumber \\
&=& 2 \langle \lambda \rangle t^{2H} \,, \quad
{\rm with} \quad
\langle \lambda \rangle 
=\int_0^\infty \lambda f(\lambda) \, d\lambda
= \lambda_0 \,
\frac{\Gamma[(\nu+1)/\rho]}{\Gamma(\nu/\rho)} \,.
\end{eqnarray}
Plots of the molecules' PDF (\ref{PDFPGf}) are shown in
Figure \ref{fig:plotPDF}.

Moreover, we consider now the Mellin transform pair \cite{marichev-1983}
\be
\mathcal{M}_r\{\varphi;s\}=\int_0^\infty \varphi(r) \, r^{s-1} \, dr \,,
\ee
\be
\varphi(r)
=\frac{1}{2\pi i}\int_{\LL} \mathcal{M}_r\{\varphi;s\} \, r^{-s} \, ds \,,
\ee
where $\LL$ is a specific contour path that separates the poles
which provide through the residue theorem a power-series with 
positive exponents from those poles that provide a power-series with
negative exponents.
Thus we obtain the following Mellin--Barnes integral representation  
\be
P(x;t) =  
\frac{1}{\Gamma(\nu/\rho)}
\frac{1}{\sqrt{4\pi\lambda_{0} t^{2H}}} 
\frac{1}{2 \pi i} \int_{\LL}  
\Gamma(s) \Gamma\left(\frac{s+\nu}{\rho} - \frac{1}{2\rho}\right)
\left[\frac{x^2}{4\lambda_{0}t^{2H}}\right]^{\!-s} ds \,.
\label{MBpdf}
\ee
For mathematical convenience,
we introduce now the change of variable 
$u=x^2 / \left[4 \lambda_{0} t^{2H}\right]$ and,
from the normalization constraint 
$\displaystyle{\int P(x;t) \, dx=\int \mathcal{P}(u) \, du=1}$, we have
\be 
\mathcal{P}(u) = 
\frac{\rho}{\Gamma(\nu/\rho)}
\frac{1}{\sqrt{4\pi u}} \, Z_\rho^{\nu-\frac{1}{2}}(u) \,.
\label{PDFu}
\ee
By using formula (\ref{MBpdf}),
the representation in terms of the H-Fox function 
\cite{luchko_etal-fcaa-2013,princy-cms-2014}
of the PDF (\ref{PDFu}) results to be
\begin{equation}
\mathcal{P}(u)=  
\frac{1}{\Gamma(\nu/\rho)}\frac{1}{\sqrt{4\pi }} 
H_{0,2}^{2,0} 
\left[ u \middle\vert 
\begin{array}{c} - \\ 
(-1/2,1) \,, ((\nu-1)/\rho,1/\rho) 
\end{array} 
\right] \,,
\label{hfoxpdf}
\end{equation}
and by considering the properties of Mellin--Barnes integrals 
\cite{marichev-1983}, 
and so the properties of the H-Fox functions as well \cite{kiryakova-1994}, 
we can get from (\ref{MBpdf}) 
the following asymptotic expansions for $u \to 0$
\begin{equation}
\mathcal{P}(u) \simeq 
\frac{1}{\Gamma(\nu/\rho)}\frac{1}{\sqrt{4\pi }} \, \mathcal{O}( u^d ) \,,
\end{equation}
with $d=\min \left[-1/2, \nu - 1 \right]$,
and for $u \to +\infty$ 
\be
\mathcal{P}(u) \simeq 
\frac{1}{\Gamma(\nu/\rho)}\frac{1}{\sqrt{4\pi }} 
\, \mathcal{O}\left(u^{\frac{\nu-\rho-1}{\rho + 1}} 
\exp \left[-(\rho+1) \left(\frac{u}{\rho}\right)^{\frac{\rho}{\rho+1}} 
\right]\right) \,.
\ee

\section{Derivation of the Fokker--Planck equation}
\label{sec:FPeq}
The problem to derive the Fokker--Planck equation of (\ref{PDFPGf})
is similar to the problem to derive the relation between 
generalised diffusion equations and subordination schemes
\cite{chechkin_etal-pre-2021}. Here, 
in order to derive the Fokker--Planck equation governing
PDF (\ref{PDFPGf}), 
we first consider the following diffusion equation 
solved by the Gaussian PDF of the fBm (\ref{gaussianfbm})
\be
\frac{\partial \G}{\partial t} = 
2 \lambda H t^{2H-1} \frac{\partial^2 \G}{\partial x^2} \,,
\label{FPfbm}
\ee
and, by multiplying both sides times $f(\lambda)$ and 
integrating over $\lambda$, from formula (\ref{PDFPGf}) we have
\begin{eqnarray}
\frac{\partial}{\partial t} P(x;t,\nu,\rho) 
&=& 2H t^{2H-1} \frac{\partial^2}{\partial x^2} 
\int_0^\infty \lambda \G(x;t| \lambda) 
f(\lambda) d\lambda \nonumber \\
&=& 2H \lambda_0 t^{2H-1} \frac{\partial^2}{\partial x^2}
\left[\frac{\Gamma((\nu+1)/\rho)}{\Gamma(\nu/\rho)} 
P(x;t,\nu+1,\rho)\right] \,,
\label{DiffeqMB}
\end{eqnarray}
where the extended notation $P(x;t,\nu,\rho)$ for PDF (\ref{PDFPGf})
is used for highlighting the difference between the PDFs 
from both sides of the equation. 

By using formula (\ref{MBpdf}), we obtain 
\begin{eqnarray}
P(x;t,\nu +a,\rho) 
&=&
\displaystyle{
\frac{\Gamma(\nu/\rho)}{\Gamma((\nu+a)/\rho)}
\frac{1}{2\pi i}\int_\mathcal{L} 
\frac{\Gamma\left(\frac{\nu+a}{\rho}-\frac{s}{2H\rho}\right)}
{\Gamma\left(\frac{\nu}{\rho}-\frac{s}{2H\rho}\right)}
\mathcal{M}_t \lbrace P(x;t,\nu,\rho) ; s \rbrace  \, t^{-s} \, ds}
\nonumber \\
&=&
\frac{\Gamma(\nu/\rho)}{\Gamma((\nu+1)/\rho)}
\, _tD_{2H\rho}^{\nu / \rho - 1,a / \rho} P(x;t,\nu,\rho) \,,
\label{MKpropint_t}
\end{eqnarray}
where $_tD_\eta^{\gamma, \mu}$,
with $\mu$, $\eta > 0$ and $\gamma \in \R$, 
is the Erd{\'e}lyi--Kober
fractional operator with respect to $t$ 
\cite[formula (3.14)]{luchko_etal-fcaa-2013}:
\begin{equation}
_tD_\eta^{\gamma, \mu} \varphi(t) = 
\dfrac{1}{2 \pi i} \int_{\mathcal{L}} 
\frac{\Gamma \left( 1+\gamma+\mu -\frac{s}{\eta}\right)}
{\Gamma \left( 1+\gamma -\frac{s}{\eta}\right)} 
\mathcal{M}_t\{\varphi;s\} \, t^{-s} ds \,.
\label{mb_ek}
\end{equation}
Therefore, 
by plugging (\ref{MKpropint_t}) with $a=1$ into (\ref{DiffeqMB}),
we have that (\ref{DiffeqMB}) is indeed
a fractional diffusion equation in the Erd{\'e}lyi--Kober sense
\begin{equation}
\frac{\partial P}{\partial t} = 2H \lambda_0 t^{2H-1} 
\, _tD_{2H\rho}^{\nu/\rho-1, 1/\rho} 
\frac{\partial^2 P}{\partial x^2} \,.
\label{EKeq}
\end{equation}
In alternative to this derivation based on the
superstatistical fBm,
another statistical perspective for the emerging of 
Erd{\'e}lyi--Kober fractional calculus comes  
from generalizations of entropy 
\cite{mathai_etal-tmj-2017,mathai_etal-cm-2017,mathai_haubold-2018}.
Moreover, for general fractional differential equations
in the Erd{\'e}lyi--Kober sense, approximations and
numerical schemes are also available 
\cite{plociniczak-siamjam-2014,plociniczak_etal-na-2017,
plociniczak_etal-fcaa-2022}.

Finally, with reference to the Golding \& Cox dataset 
\cite{golding_etal-prl-2006},
since the diffusion coefficients follow a Weibull distribution
\cite{mackala_etal-pre-2019}, i.e., $\nu=\rho$,
the generalised Fokker--Planck equation (\ref{EKeq})
turns into 
\begin{equation}
\frac{\partial P}{\partial t} = 2H \lambda_0 t^{2H-1} 
\, _tD_{2H\rho}^{0,1/\rho} 
\frac{\partial^2 P}{\partial x^2} \,,
\label{FPGC}
\end{equation}
where $2H=0.70 \pm 0.07$ \cite{golding_etal-prl-2006},
$\lambda_0^{1/2}=0.06$ and $2 \rho =1.84$
\cite{mackala_etal-pre-2019}.
In fact, if the adopted superstatistical notation is
$X_t= Y \, \mathcal{B}^H_t$ \cite{mackala_etal-pre-2019},
then variable $Y$ results to be distributed
according to the Weibull distribution $W(y;\lambda_0^{1/2},2\rho)$.
This does not affect any formula, 
but it must be accounted for when a 
comparison with empirical data is performed.

As a concluding remark, we compare the Fokker--Planck equation (\ref{EKeq}),
or its special case (\ref{FPGC}), against other equations
used in anomalous diffusion.
From formula (\ref{mb_ek}) the following noteworthy identities
can be derived \cite{luchko_etal-fcaa-2013}
\be
%_tD_1^{0,\mu} \varphi(t)= %t^{-\mu} {_tD}_1^{-\mu,\mu} f(t)=
%{_tD}_{\rm RL}^{\mu} \varphi(t) \,, \quad \mu > 0 \,,
_tD_1^{0,\mu} \varphi(t)= 
t^{-\mu} {_tD}_1^{-\mu,\mu} [t^\mu \varphi(t)]
={_tD}_{\rm RL}^{\mu} [t^\mu \varphi(t)]
\,, \quad \mu > 0 
\,,
\label{EKRL}
\ee 
where $_tD_{\rm RL}^{\mu}$ is the fractional derivative
in the Riemann--Liouville sense \cite{luchko_etal-fcaa-2013},
and we observe that formula (\ref{EKRL}) highlights 
the relation between Erd{\'e}lyi--Kober
fractional equations and Riemann--Liouville 
(or Caputo with proper initial conditions)
fractional differential equations 
with time-varying coefficients 
\cite{garra_etal-jmp-2015}.
Thus, when $\nu=\rho=1/(2H)$, 
equations (\ref{EKeq}) and (\ref{FPGC}) become 
\be
\frac{\partial P}{\partial t} = 
2H \lambda_0 t^{2H-1} \, _tD_{\rm RL}^{2H} 
\left[ t^{2H} \frac{\partial^2 P}{\partial x^2} \right] \,,
\label{TFDEGC}
\ee
that is different from existing fractional diffusion models 
with time-dependent diffusion coefficient 
\cite{fa_etal-pre-2005,garra_etal-jmp-2015,costa_etal-rmp-2021,
le-atnaa-2021}
and, moreover, it cannot be reduced further up to 
the so-called time-fractional diffusion equation 
\be
\frac{\partial \varphi}{\partial t} = 
\lambda_0 \, _tD_{\rm RL}^{1-2H} 
\frac{\partial^2 \varphi}{\partial x^2} \,.
\label{TFDE}
\ee
Equation (\ref{TFDE}) is
the fractional Fokker--Planck equation, for example,
of a continuous-time random walk (CTRW)  
with a fat-tailed distribution of waiting times,
see, e.g., \cite{metzler_etal-pr-2000},
but also for the intermediate asymptotic regime of a CTRW with
two Markovian hopping-trap mechanisms \cite{vitali_etal-jpa-2022}.
Such CTRW models have been used 
for modelling anomalous diffusion in living systems, see, e.g., 
subdiffusion of
lipid granules in living fission yeast cells \cite{jeon_etal-prl-2011},
and also, see, e.g.,  
references \cite{he_etal-prl-2008,lubelski_etal-prl-2008,neusius_etal-pre-2009},
for explaining some statistical features that appear also
in the Golding \& Cox dataset \cite{golding_etal-prl-2006}.

The failure of the CTRW approach for modelling 
some common features of anomalous diffusion in living systems
was already pointed-out 
\cite{manzo_etal-prx-2015,magdziarz_etal-prl-2009,magdziarz_etal-pre-2010}
and an alternative and promising approach seemed to be the ggBm
\cite{molina_etal-pre-2016,mackala_etal-pre-2019}.
As a matter of fact, 
the generalised Fokker--Planck equation for the ggBm emerged to be,
as well, a fractional diffusion equation in
the Erd{\'e}lyi--Kober sense as follows \cite{pagnini-fcaa-2012}:
\begin{equation}
\frac{\partial \varphi}{\partial t} = 
\lambda_0 \, 2H \rho \, t^{2H-1} 
\, _tD_{2H\rho}^{1/\rho-1,1-1/\rho} 
\frac{\partial^2 \varphi}{\partial x^2} \,,
\label{ggBmEKeq}
\end{equation}
but it is not related neither to (\ref{EKeq}) nor to (\ref{FPGC})
in anyone of their special cases. On the contrary, 
from (\ref{EKRL}) with $\rho=1/(2H)$, 
equation (\ref{ggBmEKeq}) reduces to (\ref{TFDE}), 
such that the governing equation of the CTRW 
is indeed a special case of the governing equation of the ggBm.

\begin{figure}[!h]
\centering\includegraphics[width=\textwidth]{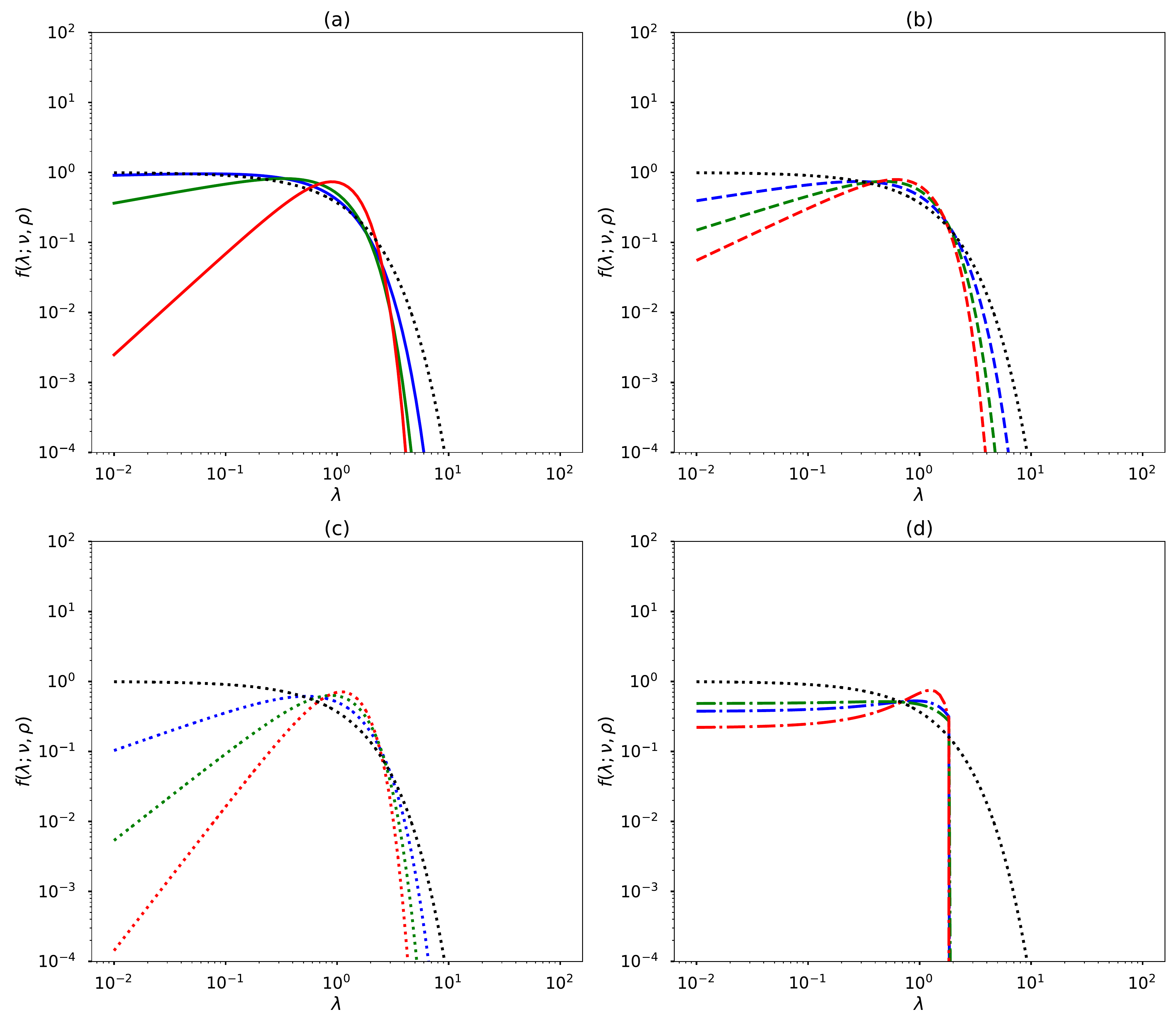}
\caption{Comparison between the generalised Gamma distribution of
the diffusion coefficients $f(\lambda)$ (\ref{gengamma}) 
and the M-function distribution $f_{\rm ggBm}(\lambda)$
(\ref{Mfunction}) for different values of the parameters.
Panel a): the generalised Gamma distribution 
[$\nu=1.25$ and $\rho=1.5$ (solid blue line),
$\nu=1.5$ and $\rho=1.75$ (solid green line),
$\nu=1.75$ and $\rho=1.25$ (solid red line)].
Panel b): the Weibull distribution 
[$\nu=\rho=1.25$ (dashed blue line),
$\nu=\rho=1.5$ (dashed green line),
$\nu=\rho=1.75$ (dashed red line)].
Panel c): the Gamma distribution ($\rho=1$)
[$\nu=1.25$ (dotted blue line),
$\nu=1.5$ (dotted green line),
$\nu=1.75$ (dotted red line)].
In each panel also 
the exponential distribution is displayed
[$\rho=\nu=1$ (dotted black line)].
In all the plots $\lambda_0=1$.
}
\label{fig:flambda}
\end{figure}

\begin{figure}[!h]
\centering\includegraphics[width=\textwidth]{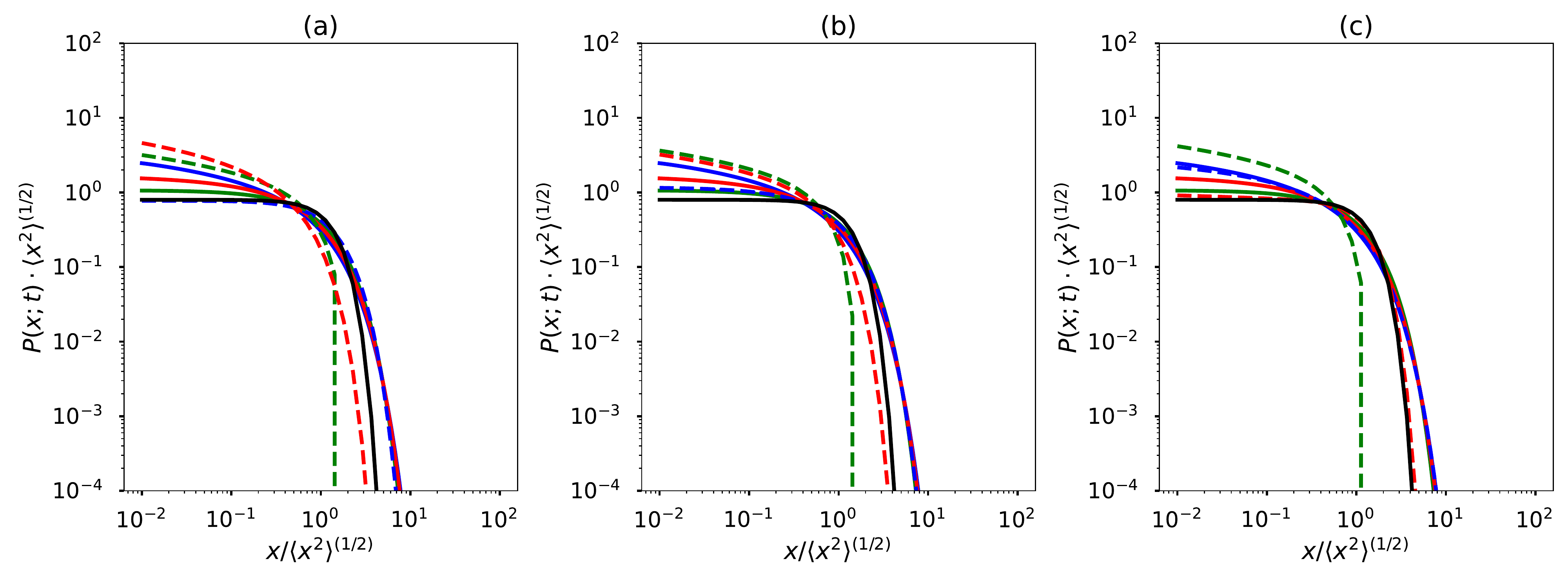}
\caption{
Comparison among the fundamental solutions 
(\ref{PDFPGf}), (\ref{PggBm}) and (\ref{PCTRW}) 
i.e.,  
$\langle x^2 \rangle^{1/2} P(x;t) \, vs \, x/\langle x^2 \rangle^{1/2}$,
in lin-log scale of the  
the corresponding Fokker--Planck equations 
(\ref{EKeq}, \ref{FPGC}, \ref{TFDEGC}), 
(\ref{ggBmEKeq}) and (\ref{TFDE}), respectively,
with the following values of the parameters:
(\ref{PDFPGf}) with $\rho=1.25$ and $\nu=1.5$ (solid blue line);
(\ref{PDFPGf}) with $\rho=\nu=1.25$ (solid green line);
(\ref{PDFPGf}) with $\rho=\nu=1.5$ (solid red line);
(\ref{PDFPGf}) with $\rho=\nu=1/(2H)$ (dashed blue line);
(\ref{PggBm}) with $\rho=X1$ (dashed green line)
and (\ref{PCTRW}) (dashed red line).
In all the plots $\lambda_0=1$.
The reference with the Gaussian density is also displayed (solid black line).
Panel a) $H=0.25$, panel b) $H=0.35$, panel c) $H=0.45$. 
}
\label{fig:PDFs}
\end{figure}

The fundamental solution of (\ref{ggBmEKeq}) is
provided by plugging in (\ref{PDFPGf}),
in place of the generalised Gamma distribution (\ref{gengamma}),
the following population of diffusion coefficients 
\be
f_{\rm ggBm}(\lambda)
=\frac{1}{\lambda_0} M_{1/\rho}\left(\frac{\lambda}{\lambda_0}\right)
=\frac{1}{\lambda_0} 
\sum_{n=0}^\infty \frac{(-\lambda/\lambda_0)^n}
{n!\Gamma[-n/\rho + (1-1/\rho)]} \,,
\label{Mfunction}
\ee
with $0 < 1/\rho < 1$, where $M_\beta(z)$ is the M-Wright Mainardi
function \cite{mainardi_etal-ijde-2010,pagnini-fcaa-2013} 
and, as a matter of fact, 
the fundamental solution of (\ref{ggBmEKeq}) here denoted by
$P_{\rm ggBm}(x;t)$ is also an M-function
\cite{mura_etal-jpa-2008,pagnini-fcaa-2012}, i.e.,
\be
P_{\rm ggBm}(x;t)= 
\frac{1}{2} 
\frac{1}{\sqrt{\lambda_0 t^{2H}}}
M_{1/(2\rho)}\left(
\frac{|x|}{\sqrt{\lambda_0 t^{2H}}}
\right) \,,
\label{PggBm}
\ee 
and then by setting $\rho=1/(2H)$ 
we have also the fundamental solution of (\ref{TFDE})
that is here denoted by $P_{\rm CTRW}(x;t)$, i.e.,
\be
P_{\rm CTRW}(x;t)= 
\frac{1}{2} 
\frac{1}{\sqrt{\lambda_0 t^{2H}}}
M_{H}\left(
\frac{|x|}{\sqrt{\lambda_0 t^{2H}}}
\right) \,.
\label{PCTRW}
\ee 
The comparison between the generalised Gamma distribution of
the diffusion coefficients $f(\lambda)$ (\ref{gengamma}) 
and the M-function distribution $f_{\rm ggBm}(\lambda)$
(\ref{Mfunction}) is shown in Figure \ref{fig:flambda}. 
Moreover, the comparison among the fundamental solutions 
(\ref{PDFPGf}), (\ref{PggBm}) and (\ref{PCTRW}) of the  
the corresponding Fokker--Planck equations 
(\ref{EKeq}, \ref{FPGC}, \ref{TFDEGC}), 
(\ref{ggBmEKeq}) and (\ref{TFDE}), respectively, 
is shown in Figure \ref{fig:PDFs} together with 
the Gaussian distribution.

\section{Discussion}
\label{sec:discussion}
In spite of the fact that knowing the solution is the 
most desiderable observable, from the equation we can indeed obtain
remarkable further information on the process.
These information can be physical, mathematical and also
useful for applications.

In fact, from the physical point-of-view
we have already observed that 
the heterogeneity of the diffusion coefficients
is the cause of the weak ergodicity breaking but from the 
Fokker--Planck equation we have also information about relation between
physical quantities like, for example, the formula of the flux
$q(x,t)$, i.e.,
\be
\frac{\partial P}{\partial t} = - \frac{\partial q}{\partial x} \,,
\ee 
that, in the cosidered case, from the Fokker--Planck equation (\ref{EKeq})
it results \cite{luchko_etal-fcaa-2013}
\begin{eqnarray}
q(x,t)
&=& - 2H \lambda_0 \, t^{2H-1} 
\, _tD_{2H\rho}^{\nu/\rho-1, 1/\rho}  
\frac{\partial P}{\partial x} \nonumber \\
&=& - \frac{4 H^2 \, \rho \lambda_0 \, t^{2H-1}}{\Gamma(n-1/\rho)}
\times \nonumber \\
& & 
\frac{\partial}{\partial x} 
\prod_{j=1}^n \left[
\frac{\nu}{\rho} -1 +j + \frac{t}{2H\rho}\frac{\partial}{\partial t}
\right]
\left\{
\frac{
\displaystyle{\int_0^t 
\frac{(t^{2H\rho}-\tau^{2H\rho})^{n-1-1/\rho}}
{\tau^{1-2H(\nu+1)}} P(x;\tau) \, d\tau}}
{t^{2H[\nu + \rho(n-1)]}} \right\} \,,
\label{flux}
\end{eqnarray} 
with $n-1 < 1/\rho \le 1$,
that is not proportional to the gradient as in standard diffusion
and we have indeed that the heterogeneity is the cause
of the emergence of a memory kernel. 
This memory kernel is determined by the specific distribution
of the diffusion coefficients $f(\lambda)$.
Moreover, because of this memory kernel it follows that the 
resulting evolution equation is a fractional differential equation, 
and a special
connection exists between the generalised Gamma distribution and
the Erd{\'e}lyi--Kober fractional operators,
see \cite{mathai_etal-tmj-2017,mathai_etal-cm-2017,mathai_haubold-2018}.

Moreover, from the governing equation, 
we can see if there are forcing terms and their origin
and, actually, we have from (\ref{EKeq}) that the heterogeneity 
of the diffusion coefficient causes a flux with memory but it
does not cause the emerging of any apparent forcing:
in fact the resulting governing equation does not include any terms
like $\partial F/\partial x$ and preserves indeed the form of 
the free-particle diffusion equation with the feature
of a time-dependent effective diffusion coefficient. 
In particular, with respect to space,
this heterogenity does not cause any effect 
and the operator in space remains 
the second derivative $\partial^2/\partial x^2$ as in 
standard diffusion.

Furthermore, 
the derivation of Fokker--Planck equation (\ref{EKeq}) embodies 
as a matter of fact a derivation on physical ground 
of a fractional equation by avoiding the replacement of
the operators. Actually, this derivation,
together with the derivation of the governing equation 
of the ggBm \cite{pagnini-fcaa-2012}, 
is an alternative derivation with respect to that 
on statistical grounds discussed in literature
\cite{mathai_etal-tmj-2017,mathai_etal-cm-2017,mathai_haubold-2018}
of fractional differential equations in the 
Erd{\'e}lyi--Kober sense: 
this physical significance of fractional equations let indeed
to overpass questioned issues in fractional-dynamic generalizations
and some related constraints that have to be checked \cite{tarasov-m-2019}.  

From the mathematical point-of-view,
we observe indeed that the process turns into a process
governed by an integro-differential equation but
it remains linear and parabolic, 
namely the emergence of a memory kernel is not the cause,
for example, of the emergence of a second derivative in time,
i.e, $\partial^2 P/\partial t^2$, or a finite front velocity,
by keeping the main characteristics of the standard diffusion.

To conclude this section about the importance to know the
governing equation of a process 
rather than the solution in a special case,
we report that,
under the practical point of view and in the direction of applications,
to know the equation allows for studying, at least numerically, 
the solution with particular initial and boundary conditions 
and also with real settings in complex and multidimensional domains,
as well as, in presence of external forcings.

\section{Conclusions and future perspectives}
\label{sec:conclusions}
In this paper we focused on the signature of anomalous diffusion 
as measured in many biological systems and in particular 
as it is measured in the motion of passive tracers inside cytoplasm as
the mRNA molecules inside live E. coli cells.
From the wide literature, we paied special attention
to the dataset by Golding \& Cox \cite{golding_etal-prl-2006}
that has been largely analysed, see, e.g.,
\cite{golding_etal-prl-2006,
magdziarz_etal-prl-2009,
magdziarz_etal-pre-2010,
burnecki_etal-pre-2010,
weron_etal-prl-2010,
magdziarz_etal-pre-2011,
mackala_etal-pre-2019,
janczura_etal-csf-2022}.
Among the many distinctive features, 
the most important for the present study is the determination of
the single-molecule trajectory as a fBm,
see, e.g., \cite{
magdziarz_etal-prl-2009,
szymanski_etal-prl-2009,
magdziarz_etal-pre-2010,
weiss-pre-2013,krapf_etal-prx-2019,
sabri_etal-prl-2020,
han_etal-elife-2020,itto_etal-jrsi-2021,korabel_etal-e-2021},
and also the heterogeneity of the diffusion coefficient
according to a Weibull-type distribution
that emerges from the data analysis when  
the observed anomalous behaviour is reproduced through  
a fBm as the underlying stochastic process
\cite{mackala_etal-pre-2019}.

Following these experimental evidences, 
we slightly generalised the distribution of the diffusion coefficients
by using a generalised Gamma distribution in analogy with
other similar analysis 
\cite{hapca_etal-jrsi-2009,petrovskii_etal-an-2009,manzo_etal-prx-2015}
and we showed that the PDF of the mRNA molecule displacement is
related with the Kr{\"a}tzel function and we
derived the corresponding Fokker--Planck equation. 
The governing equation results to be a fractional diffusion equation 
in the Erd{\'e}lyi--Kober sense with a time-dependent effective
diffusion coefficient, 
which cannot be reduced to the governing equations of existing
literature models as the CTRW or the ggBm, 
neither to other models governed by a fractional differential equation
with a time-dependent diffusion coefficient.

This last fact is indeed a two-fold findings 
for motivating future investigations on this model setting.
In fact, 
the derived equation (\ref{EKeq}) pushes the research
towards the mathematical analysis of a novel family of equations 
as well as towards the development of solid and
reliable numerical schemes for studying more realistic systems
in multidimensional domains with 
real geometries and general initial and boundary condition, 
but also towards the development of statistical methods and
tools for proper taking into account a distribution of diffusion
coefficients that can lead to different modelling approaches 
as those based on an heterogeneous ensemble of particles,
e.g., the over- and under-damped ggBm 
\cite{mura_etal-jpa-2008,molina_etal-pre-2016,
vitali_etal-jrsi-2018,sposini_etal-njp-2018,mackala_etal-pre-2019,
sliusarenko_etal-jpa-2019,dossantos_etal-p-2020,chen_etal-njp-2021}, or 
those based on diffusion in inhomogeneous random environments,
e.g., the diffusing diffusivity approach 
\cite{chubynsky_etal-prl-2014,chechkin_etal-prx-2017,sposini_etal-njp-2018,
dossantos_etal-csf-2021a,dossantos_etal-csf-2021b}.

Finally, we highlight that the present framework 
does not include, yet, two quite general and well established features 
of anomalous diffusion: the Brownian yet non-Gaussian regime 
\cite{chubynsky_etal-prl-2014,chechkin_etal-prx-2017,metzler-bj-2017,
sposini_etal-njp-2018,postnikov_etal-njp-2020,
wang_etal-pa-2021,alban_etal-m-2022}
and the anomalous-to-normal transition 
\cite{saichev_etal-jetp-2004,
metzler-bj-2017,molina_etal-njp-2018,sliusarenko_etal-jpa-2019}
that are indeed part of the paradigm of anomalous diffusion 
\cite{vitali_etal-jpa-2022} and observed 
both in the continuous-space setting of the diffusion diffusivity models 
\cite{chechkin_etal-prx-2017,sposini_etal-njp-2018} 
and in the discrete-space setting of random walks \cite{saichev_etal-jetp-2004}
even with solely
two Markovian hopping-trap mechanisms \cite{vitali_etal-jpa-2022}.
Thus, these last embody the future developments of research 
on the superstatistical fBm. 

\section*{Acknowledgements}
This research is supported by the Basque Government through 
the BERC 2018--2021 and 2022--2025 programs and by the Ministry of Science, 
Innovation and Universities: BCAM Severo Ochoa accreditation SEV-2017-0718.
The authors acknowledge Prof. F. Mainardi for useful hints and 
support received on special functions and anomalous diffusion,
and Prof. G. Castellani for contributing to build up 
this collaboration.
The research was carried out under the auspices of INDAM-GNFM 
(the National Group of Mathematical Physics of 
the Italian National Institute of High Mathematics).


\begin{thebibliography}{99}
\bibitem{barkai_etal-pt-2012}
Barkai E, Garini Y, Metzler R. 2012  {Strange kinetics of single molecules in
  living cells}. {\em Phys. Today} \textbf{65}, 29.

\bibitem{hofling_etal-rpp-2013}
H{\"{o}}fling F, Franosch T. 2013  Anomalous transport in the crowded world of
  biological cells. {\em Rep. Prog. Phys.} \textbf{76}, 046602.

\bibitem{manzo_etal-rpp-2015}
Manzo C, Garcia-Parajo MF. 2015  A review of progress in single particle
  tracking: from methods to biophysical insights. {\em Rep. Progr. Phys.}
  \textbf{78}, 124601.

\bibitem{golding_etal-prl-2006}
Golding I, Cox EC. 2006  Physical nature of bacterial cytoplasm. {\em Phys.
  Rev. Lett.} \textbf{96}, 098102.

\bibitem{magdziarz_etal-prl-2009}
Magdziarz M, Weron A, Burnecki K, Klafter J. 2009  Fractional {Brownian} motion
  versus the {Continuous-Time Random Walk}: {A} simple test for subdiffusive
  dynamics. {\em Phys. Rev. Lett.} \textbf{103}, 180602.

\bibitem{magdziarz_etal-pre-2010}
Magdziarz M, Klafter J. 2010  Detecting origins of subdiffusion: p-variation
  test for confined systems. {\em Phys. Rev. E} \textbf{82}, 011129.

\bibitem{burnecki_etal-pre-2010}
Burnecki K, Weron A. 2010  Fractional {L\'evy} stable motion can model
  subdiffusive dynamics. {\em Phys. Rev. E} \textbf{82}, 021130.

\bibitem{weron_etal-prl-2010}
Weron A, Magdziarz M. 2010  Generalization of the {Khinchin} Theorem to
  {L\'evy} Flights. {\em Phys. Rev. Lett.} \textbf{105}, 260603.

\bibitem{magdziarz_etal-pre-2011}
Magdziarz M, Weron A. 2011  Anomalous diffusion: {Testing} ergodicity breaking
  in experimental data. {\em Phys. Rev. E} \textbf{84}, 051138.

\bibitem{mackala_etal-pre-2019}
Ma{\'c}ka{\l{}}a A, Magdziarz M. 2019  Statistical analysis of superstatistical
  fractional {Brownian} motion and applications. {\em Phys. Rev. E}
  \textbf{99}, 012143.

\bibitem{janczura_etal-csf-2022}
Janczura J, Burnecki K, Muszkieta M, Stanislavsky A, Weron A. 2022
  Classification of random trajectories based on the fractional {L\'evy} stable
  motion. {\em Chaos Solitons Fract.} \textbf{154}, 111606.

\bibitem{tolicnorrelykke_etal-prl-2004}
Toli\'c-N{\o}rrelykke IM, Munteanu EL, Thon G, Odderhede L, Berg-S{\o}rensen K.
  2004  Anomalous diffusion in living yeast cells. {\em Phys. Rev. Lett.}
  \textbf{93}, 078102.

\bibitem{bronstein_etal-prl-2009}
Bronstein I, Israel Y, Kepten E, Mai S, Shav-Tal Y, Barkai E, Garini Y. 2009
  Transient Anomalous Diffusion of Telomeres in the Nucleus of Mammalian Cells.
  {\em Phys. Rev. Lett.} \textbf{103}, 018102.

\bibitem{weigel_etal-pnas-2011}
Weigel AV, Simon B, Tamkun MM, Krapf D. 2011  {Ergodic and nonergodic processes
  coexist in the plasma membrane as observed by single-molecule tracking.}.
  {\em Proc. Natl. Acad. Sci. USA} \textbf{108}, 6438--43.

\bibitem{tabei_etal-pnas-2013}
Tabei SMA, Burov S, Kim HY, Kuznetsov A, Huynh T, Jureller J, Philipson LH,
  Dinner AR, Scherer NF. 2013  Intracellular transport of insulin granules is a
  subordinated random walk. {\em Proc. Natl. Acad. Sci. USA} \textbf{110},
  4911--4916.

\bibitem{regner_etal-bj-2013}
Regner BM, Vu\v{c}ini\'{c} D, Domnisoru C, Bartol TM, Hetzer MW, Tartakovsky
  DM, Sejnowski TJ. 2013  Anomalous diffusion of single particles in cytoplasm.
  {\em Biophys. J.} \textbf{104}, 1652--1660.

\bibitem{jeon_etal-prl-2011}
Jeon J, Tejedor V, Burov S, Barkai E, {Selhuber--Unkel} C, Berg-S{\o}rensen K,
  Oddershede L, Metzler R. 2011  In Vivo Anomalous Diffusion and Weak
  Ergodicity Breaking of Lipid Granules. {\em Phys. Rev. Lett.} \textbf{106},
  048103.

\bibitem{manzo_etal-prx-2015}
Manzo C, {Torreno--Pina} JA, Massignan P, Lapeyre GJ, Lewenstein M,
  {Garcia--Parajo} MF. 2015  {Weak Ergodicity Breaking of Receptor Motion in
  Living Cells Stemming from Random Diffusivity}. {\em Phys. Rev. X}
  \textbf{5}, 011021.

\bibitem{szymanski_etal-prl-2009}
Szymanski J, Weiss M. 2009  Elucidating the Origin of Anomalous Diffusion in
  Crowded Fluids. {\em Phys. Rev. Lett.} \textbf{103}, 038102.

\bibitem{weiss-pre-2013}
Weiss M. 2013  Single-particle tracking data reveal anticorrelated fractional
  {Brownian} motion in crowded fluids. {\em Phys. Rev. E} \textbf{88},
  010101(R).

\bibitem{krapf_etal-prx-2019}
Krapf D, Lukat N, Marinari E, Metzler R, Oshanin G, {Selhuber--Unkel} C,
  Squarcini A, Stadler L, Weiss M, Xu X. 2019  Spectral Content of a Single
  {non-Brownian} Trajectory. {\em Phys. Rev. X} \textbf{9}, 011019.

\bibitem{sabri_etal-prl-2020}
Sabri A, Xu X, Krapf D, Weiss M. 2020  Elucidating the Origin of Heterogeneous
  Anomalous Diffusion in the Cytoplasm of Mammalian Cells. {\em Phys. Rev.
  Lett.} \textbf{125}, 058101.

\bibitem{han_etal-elife-2020}
Han D, Korabel N, Chen R, Johnston M, Gavrilova A, Allan VJ, Fedotov S, Waigh
  TA. 2020  Deciphering anomalous heterogeneous intracellular transport with
  neural networks. {\em eLife} \textbf{9}, e52224.

\bibitem{itto_etal-jrsi-2021}
Itto Y, Beck C. 2021  Superstatistical modelling of protein diffusion dynamics
  in bacteria. {\em J. R. Soc. Interface} \textbf{18}, 20200927.

\bibitem{korabel_etal-e-2021}
Korabel N, Han D, Taloni A, Pagnini G, Fedotov S, Allan V, Waigh TA. 2021
  Local Analysis of Heterogeneous Intracellular Transport: Slow and Fast Moving
  Endosomes. {\em Entropy} \textbf{23}, 958.

\bibitem{hapca_etal-jrsi-2009}
Hapca S, Crawford JW, Young IM. 2009  Anomalous diffusion of heterogeneous
  populations characterized by normal diffusion at the individual level. {\em
  J. R. Soc. Interface} \textbf{6}, 111--122.

\bibitem{petrovskii_etal-an-2009}
Petrovskii S, Morozov A. 2009  Dispersal in a statistically structured
  population: fat tails revisited. {\em Am. Nat.} \textbf{173}, 278--289.

\bibitem{molina_etal-pre-2016}
Molina-Garc\'ia D, {Minh Pham} T, Paradisi P, Manzo C, Pagnini G. 2016
  Fractional kinetics emerging from ergodicity breaking in random media. {\em
  Phys. Rev. E} \textbf{94}, 052147.

\bibitem{mura_etal-jpa-2008}
Mura A, Pagnini G. 2008  Characterizations and simulations of a class of
  stochastic processes to model anomalous diffusion. {\em J. Phys. A: Math.
  Theor.} \textbf{41}, 285003.

\bibitem{grothaus_etal-jfa-2015}
Grothaus M, Jahnert F, Riemann F, {da Silva} JL. 2015  {Mittag--Leffler
  analysis I: Construction} and characterization. {\em J. Funct. Anal.}
  \textbf{268}, 1876--1903.

\bibitem{grothaus_etal-jfa-2016}
Grothaus M, Jahnert F. 2016  {Mittag--Leffler analysis II: Application} to the
  fractional heat equation. {\em J. Funct. Anal.} \textbf{270}, 2732--2768.

\bibitem{dasilva_etal-s-2015}
{da Silva} JL, Erraoui M. 2015  Generalized grey {Brownian} motion local time:
  existence and weak approximation. {\em Stochastics} \textbf{87}, 347--361.

\bibitem{bender_etal-fcaa-2022a}
Bender C, Butko YA. 2022  Stochastic solutions of generalized time-fractional
  evolution equations. {\em Fract. Calc. Appl. Anal.} \textbf{25}, 488--519.

\bibitem{bender_etal-fcaa-2022b}
Bender C, Bormann M, Butko YA. 2022  Subordination principle and {Feynman--Kac}
  formulae for generalized time-fractional evolution equations. {\em Fract.
  Calc. Appl. Anal.}
In press: https://doi.org/10.1007/s13540-022-00082-8.

\bibitem{vitali_etal-jrsi-2018}
Vitali S, Sposini V, Sliusarenko O, Paradisi P, Castellani G, Pagnini G. 2018
  Langevin equation in complex media and anomalous diffusion. {\em J. R. Soc.
  Interface} \textbf{15}, 20180282.

\bibitem{sliusarenko_etal-jpa-2019}
Sliusarenko O, Vitali S, Sposini V, Paradisi P, Chechkin A, Castellani G,
  Pagnini G. 2019  Finite-energy {L\'evy}-type motion through heterogeneous
  ensemble of {Brownian} particles. {\em J. Phys. A} \textbf{52}, 095601.

\bibitem{chen_etal-njp-2021}
Chen Y, Wang X. 2021  Novel anomalous diffusion phenomena of underdamped
  {Langevin} equation with random parameters. {\em New J. Phys.} \textbf{23},
  123024.

\bibitem{vitali_etal-m-2019}
Vitali S, Budimir I, Runfola C, Castellani G. 2019  The Role of the Central
  Limit Theorem in the Heterogeneous Ensemble of Brownian Particles Approach.
  {\em Mathematics} \textbf{7}, 1145.

\bibitem{kalwarczyk_etal-jpcb-2017}
Kalwarczyk T, Kwapiszewska K, Szczepanski K, Sozanski K, Szymanski J, Michalska
  B, {Patalas--Krawczyk} P, Duszynski J, Holyst R. 2017  Apparent Anomalous
  Diffusion in the Cytoplasm of Human Cells: The Effect of Probes'
  Polydispersity. {\em J. Phys. Chem. B} \textbf{121}, 9831--9837.

\bibitem{princy-cms-2014}
Princy T. 2014  Kr{\"a}tzel Function and Related Statistical Distributions.
  {\em Commun. Math. Stat.} \textbf{2}, 413--429.

\bibitem{luchko_etal-fcaa-2013}
Luchko Y, Kiryakova V. 2013  The {Mellin} integral transform in {Fractional
  Calculus}. {\em Fract. Calc. Appl. Anal.} \textbf{16}, 405--430.

\bibitem{mathai_haubold-2018}
Mathai AM, Haubold HJ. 2018 {\em {Erd{\'e}lyi--Kober} Fractional Calculus. From
  a Statistical Perspective, Inspired by Solar Neutrino Physics}.
Singapore: Springer Nature.

\bibitem{weiss_etal-bj-2004}
Weiss M, Elsner M, Kartberg F, Nilsson T. 2004  Anomalous subdiffusion is a
  measure for cytoplasmic crowding in living cells. {\em Biophys. J.}
  \textbf{87}, 3518--3524.

\bibitem{banks_etal-bj-2005}
Banks DS, Fradin C. 2005  Anomalous diffusion of proteins due to molecular
  crowding. {\em Biophys. J.} \textbf{89}, 2960--2971.

\bibitem{ernst_etal-sm-2012}
Ernst D, Hellmann M, K{\"{o}}hler J, Weiss M. 2012  {Fractional Brownian
  motion} in crowded fluids. {\em Soft Matter} \textbf{8}, 4886--4889.

\bibitem{sadoon_etal-pre-2018}
Sadoon AA, Wang Y. 2018  {Anomalous, non-Gaussian, viscoelastic, and
  age-dependent dynamics of histonelike nucleoid-structuring proteins in live
  Escherichia coli}. {\em Phys. Rev. E} \textbf{98}, 042411.

\bibitem{beck-pa-2006}
Beck C. 2006  Stretched exponentials from superstatistics. {\em Physica A}
  \textbf{365}.

\bibitem{sposini_etal-njp-2018}
Sposini V, Chechkin AV, Seno F, Pagnini G, Metzler R. 2018  Random diffusivity
  from stochastic equations: {Comparison} of two models for {Brownian} yet
  {non-Gaussian} diffusion. {\em New J. Phys.} \textbf{20}, 043044.

\bibitem{postnikov_etal-njp-2020}
Postnikov EB, Chechkin A, Sokolov IM. 2020  Brownian yet {non-Gaussian}
  diffusion in heterogeneous media: from superstatistics to homogenization.
  {\em New J. Phys.} \textbf{22}, 063046.

\bibitem{marichev-1983}
Marichev OI. 1983 {\em Handbook of Integral Transforms of Higher Trascendental
  Functions, Theory and Algorithmic Tables}.
Ellis Horwood, Chichester.

\bibitem{kiryakova-1994}
Kiryakova V. 1994 {\em Generalized Fractional Calculus and Applications}.
Harlow/New York: Longman Scientific \& Technical/John Wiley \& Sons Inc.
[Pitman Research Notes in Mathematics, vol. 301].

\bibitem{chechkin_etal-pre-2021}
Chechkin A, Sokolov IM. 2021  Relation between generalized diffusion equations
  and subordination schemes. {\em Phys. Rev. E} \textbf{103}, 032133.

\bibitem{mathai_etal-tmj-2017}
Mathai AM, Haubold HJ. 2017a  {Erd{\'e}lyi--Kober} fractional integral
  operators from a statistical perspective {(I)}. {\em Tbil. Math. J.}
  \textbf{10}, 145--159.

\bibitem{mathai_etal-cm-2017}
Mathai AM, Haubold HJ. 2017b  {Erd{\'e}lyi--Kober} fractional integral
  operators from a statistical perspective {-II}. {\em Cogent Math.}
  \textbf{4}, 1309769.

\bibitem{plociniczak-siamjam-2014}
P{\l{}}ociniczak {\L{}}. 2014  Approximation of the {Erd{\'e}lyi--Kober}
  operator with application to the time-fractional porous medium equation. {\em
  SIAM J. Appl. Math.} pp. 1219--1237.

\bibitem{plociniczak_etal-na-2017}
P{\l{}}ociniczak {\L{}}, Sobieszek S. 2017  Numerical schemes for
  integro-differential equations with {Erd{\'e}lyi--Kober} fractional operator.
  {\em Numer. Algor.} \textbf{76}, 125--150.

\bibitem{plociniczak_etal-fcaa-2022}
P{\l{}}ociniczak {\L{}}, {\'S}wita{\l{}}a M. 2022  Numerical scheme for
  {Erd{\'e}lyi--Kober} fractional diffusion equation using {Galerkin--Hermite}
  method. {\em Fract. Calc. Appl. Anal.} \textbf{25}, 1651--1687.

\bibitem{garra_etal-jmp-2015}
Garra R, Orsingher E, Polito F. 2015  Fractional diffusions with time-varying
  coefficients. {\em J. Math. Phys.} \textbf{56}, 093301.

\bibitem{fa_etal-pre-2005}
Fa KS, Lenzi EK. 2005  Time-fractional diffusion equation with time dependent
  diffusion coefficient. {\em Phys. Rev. E} \textbf{72}, 011107.

\bibitem{costa_etal-rmp-2021}
Costa FS, {Capelas de Oliveira} E, Plata ARG. 2021  Fractional diffusion with
  time-dependent diffusion coefficient. {\em Rep. Math. Phys.} \textbf{87},
  59--79.

\bibitem{le-atnaa-2021}
Le DL. 2021  Note on a time fractional diffusion equation with time dependent
  variables coefficients. {\em Adv. Theory Nonlinear Anal. Appl.} \textbf{5},
  600--610.

\bibitem{metzler_etal-pr-2000}
Metzler R, Klafter J. 2000  The random walk's guide to anomalous diffusion: a
  fractional dynamics approach. {\em Phys. Rep.} \textbf{339}, 1--77.

\bibitem{vitali_etal-jpa-2022}
Vitali S, Paradisi P, Pagnini G. 2022  Anomalous diffusion originated by two
  {Markovian} hopping-trap mechanisms. {\em J. Phys. A: Math. Theor.}
  \textbf{55}, 224012.

\bibitem{he_etal-prl-2008}
He Y, Burov S, Metzler R, Barkai E. 2008  Random time-scale invariant diffusion
  and transport coefficients. {\em Phys. Rev. Lett.} \textbf{101}, 058101.

\bibitem{lubelski_etal-prl-2008}
Lubelski A, Sokolov IM, Klafter J. 2008  Nonergodicity mimics inhomogeneity in
  single particle tracking. {\em Phys. Rev. Lett.} \textbf{100}, 250602.

\bibitem{neusius_etal-pre-2009}
Neusius T, Sokolov IM, Smith JC. 2009  Subdiffusion in time-averaged, confined
  random walks. {\em Phys. Rev. E} \textbf{80}, 011109.

\bibitem{pagnini-fcaa-2012}
Pagnini G. 2012  Erd\'elyi--{Kober} fractional diffusion. {\em Fract. Calc.
  Appl. Anal.} \textbf{15}, 117--127.

\bibitem{mainardi_etal-ijde-2010}
Mainardi F, Mura A, Pagnini G. 2010  The {M-Wright} function in time-fractional
  diffusion processes: {A} tutorial survey. {\em Int. J. Differ. Equations}
  \textbf{2010}, 104505.

\bibitem{pagnini-fcaa-2013}
Pagnini G. 2013  The {M-Wright} function as a generalization of the {Gaussian}
  density for fractional diffusion processes. {\em Fract. Calc. Appl. Anal.}
  \textbf{16}, 436--453.

\bibitem{tarasov-m-2019}
Tarasov VE. 2019  Rules for fractional-dynamic generalizations: {Difficulties}
  of constructing fractional dynamic models. {\em Mathematics} \textbf{7}, 554.

\bibitem{dossantos_etal-p-2020}
{dos Santos} MAF, {Menon Junior} L. 2020  {Log-Normal} superstatistics for
  {Brownian} Particles in a Heterogeneous Environment. {\em Physics}
  \textbf{2}, 571--586.

\bibitem{chubynsky_etal-prl-2014}
Chubynsky M, Slater G. 2014  Diffusing Diffusivity: A Model for Anomalous, yet
  Brownian, Diffusion. {\em Phys. Rev. Lett.} \textbf{113}, 098302.

\bibitem{chechkin_etal-prx-2017}
Chechkin AV, Seno F, Metzler R, Sokolov IM. 2017  Brownian yet {non-Gaussian}
  Diffusion: From Superstatistics to Subordination of Diffusing Diffusivities.
  {\em Phys. Rev. X} \textbf{7}, 021002.

\bibitem{dossantos_etal-csf-2021a}
{dos Santos} MAF, {Menon Junior} L. 2021  Random diffusivity models for scaled
  {Brownian motion}. {\em Chaos Solitons Fract.} \textbf{144}, 110634.

\bibitem{dossantos_etal-csf-2021b}
{dos Santos} MAF, Colombo EH, Anteneodo C. 2021  Random diffusivity scenarios
  behind anomalous {non-Gaussian} diffusion. {\em Chaos Solitons Fract.}
  \textbf{152}, 111422.

\bibitem{metzler-bj-2017}
Metzler R. 2017  Gaussianity fair: {The} riddle of anomalous yet non-{Gaussian}
  diffusion. {\em Biophys. J.} \textbf{112}, 413--415.

\bibitem{wang_etal-pa-2021}
Wang X, Chen Y. 2021  Ergodic property of {Langevin} systems with
  superstatistical, uncorrelated or correlated diffusivity. {\em Physica A}
  \textbf{577}, 126090.

\bibitem{alban_etal-m-2022}
{Alban--Chac{\'o}n} FE, Lamilla-Rubio EA, Alvarez-Alvarado MS. 2022  A novel
  physical mechanism to model {Brownian yet non-Gaussian diffusion: Theory} and
  application. {\em Materials} \textbf{15}, 5808.

\bibitem{saichev_etal-jetp-2004}
Saichev AI, Utkin SG. 2004  Random Walks with Intermediate Anomalous-Diffusion
  Asymptotics. {\em J. Exp. Theor.} \textbf{99}, 443--448.
Translated from Zhurnal {\'E}ksperimental' no{\u{\i}} i Teoretichesko{\u{\i}}
  Fiziki, Vol. 126, No. 2, 2004, pp. 502--508.

\bibitem{molina_etal-njp-2018}
Molina-Garc\'ia D, Sandev T, Safdari H, Pagnini G, Chechkin A, Metzler R. 2018
  Crossover from anomalous to normal diffusion: truncated power-law noise
  correlations and applications to dynamics in lipid bilayers. {\em New J.
  Phys.} \textbf{20}, 103027.

\end{thebibliography}
\end{document}